\documentclass[preprint]{aastex}
\usepackage[dvips]{epsfig}
\usepackage[dvips]{graphicx}
\usepackage[dvips]{color}
\usepackage[stable]{footmisc}
\usepackage{natbib}
\usepackage{amssymb}  
\begin{document}
\def\intunits{\rm s^{-1}\,sr^{-1}\,cm^{-2}}
\def\diffunits{\rm GeV s^{-1}\,sr^{-1}\,cm^{-2}}
\def\fluxunits{\rm GeV^{-1} s^{-1}\,sr^{-1}\,cm^{-2}}
\def\diffcr{\frac{dN_{\rm CR}}{dE_{\rm CR}}}
\def\diffp{\frac{dN_{p}}{dE_{p}}}
\def\diffnu{\frac{dN_{\nu}}{dE_{\nu}}}
\def\en{E_{\nu}}
\def\eg{E_{\gamma}}
\def\ep{E_{p}}
\def\ecr{E_{\rm CR}}
\def\ee{E_{e}}
\def\epb{\epsilon_{p}^{b}}
\def\enb{\epsilon_{\nu}^{b}}
\def\enbG{\epsilon_{\nu,GeV}^{b}}
\def\enbM{\epsilon_{\nu,MeV}^{b}}
\def\ens{\epsilon_{\nu}^{s}}
\def\ensG{\epsilon_{\nu,GeV}^{s}}
\def\egb{\epsilon_{\gamma}^{b}}
\def\egbM{\epsilon_{\gamma,MeV}^{b}}
\def\eauger{E_{\rm Auger}^{\min}}
\def\emincr{E_{\rm p}^{\min}}
\def\eminnu{E_{\rm \nu}^{\min}}
\def\lumi{L_{\gamma}^{52}}
\title{No observational constraints from hypothetical collisions
of hypothetical dark halo primordial black holes with Galactic
objects} \shorttitle{No constraints from PBH collisions in the
Galaxy}
\author{Marek A.~Abramowicz\altaffilmark{1,2}, Julia K.~Becker\altaffilmark{1,3},
  Peter L.~Biermann\altaffilmark{4,5,6,7,8}\\Antonella Garzilli\altaffilmark{1,9},
  Fredrik Johansson\altaffilmark{1,10} and Lei Qian\altaffilmark{11}}
\shortauthors{M.~A.~Abramowicz et al.}
\altaffiltext{1}{G\"oteborgs Universitet, Institutionen f\"or Fysik, SE-41296
G\"oteborg, Sweden}
\altaffiltext{2}{N. Copernicus Astronomical Centre, Polish Academy of
  Sciences, Bartycka 18, 00-716 Warszawa, Poland}
\altaffiltext{3}{Fakult\"at f.\ Physik \& Astronomie, Theoretische
  Physik IV, Ruhr-Univerist\"at Bochum, Bochum, Germany}
\altaffiltext{4}{Max Planck Institut f\"ur Radioastronomie, Auf dem H\"ugel 69, D-53121
  Bonn, Germany}
\altaffiltext{5}{Department of Physics and Astronomy, University of Bonn,
  Germany}
\altaffiltext{6}{Department of Physics and Astronomy, University of
  Alabama, Tuscaloosa, AL, USA}
\altaffiltext{7}{Department of Physics and Astronomy, University of Alabama at Huntsville, AL, USA}
\altaffiltext{8}{Inst.~Nucl.~Phys.~FZ, Karlsruhe Inst.~of Techn.~(KIT), Karlsruhe, Germany}
\altaffiltext{9}{SISSA, via Beirut 2-4, 34151 Trieste, Italy}
\altaffiltext{10}{Chalmers tekniska h\"ogskola, Institutionen f\"or fundamental fysik, SE-41296 G\"oteborg, Sweden}
\altaffiltext{11}{Astronomy Department, School of Physics, Peking
  University, Beijing 100871, P. R. China}
\begin{abstract}
\noindent It was suggested by several authors
that hypothetical primordial black holes (PBHs) may contribute to
the dark matter in our Galaxy. There are strong constraints based
on the Hawking evaporation that practically exclude PBHs with
masses $m_{pbh}\sim 10^{15}-10^{16}$~g and smaller as significant
contributors to the Galactic dark matter. Similarly, PBHs with
masses greater than about $10^{26}$~g are practically excluded by
the gravitational lensing observation. The mass range between
$10^{16}$~g~$<m_{pbh}<10^{26}$~g is unconstrained. In this paper,
we examine possible observational signatures in the unexplored
mass range, investigating hypothetical collisions of PBHs with
main sequence stars, red giants, white dwarfs, and neutron stars
in our Galaxy. This has previously been discussed as possibly
leading to an observable photon eruption due to shock production
during the encounter. We find that such collisions are either too
rare to be observed (if the PBH masses are typically larger than
about $10^{20}$~g), or produce too little power to be detected (if
the masses are smaller than about $10^{20}$~g).
 \end{abstract}
\parindent=0cm
\parskip=0.2cm
\section{Introduction}
 The idea of dark matter (DM) in the Universe was first advanced in the
1930s, when~\cite{zwicky1933,zwicky1937} discussed the deviation
of the masses of galaxy clusters away from expected values. The
idea was subsequently confirmed in many different
ways~\citep[e.g.][]{ostriker_peebles_yahil1974,wmap_5yr}. The
strong deviation of the rotation curves of spiral galaxies away
from what would result from just the luminous matter, has given an
important additional quantitative indication of the existence of
DM. Its presence in our Galaxy was first discussed
by~\cite{kahn_woltjer1959}. The current estimate gives
$M_{DM}=9\cdot 10^{11}\,M_{\odot}$~\citep{xue2008}, which matches
that by~\cite{kahn_woltjer1959}. Searches for non-luminous DM in
the form of brown and white dwarfs, neutron stars and black holes,
with masses above $0.1-0.2\cdot M_{\odot}$ (``MACHOS'') has shown
that these objects are far too rare to make up a significant part
of the total DM \citep{afonso2003,yoo2004}. Within the past five
years, it has even been shown by WMAP that baryonic matter only
makes up $\sim 4\%$, of the critical density of the Universe while
dark matter makes up $\sim 20\%$~\citep{wmap_5yr}. While objects
like neutron stars or white dwarfs are thereby excluded as making
up a major proportion of the DM, black holes with masses between
$10^{16}\,{\rm g} <m_{pbh}<10^{26}{\rm g}$ might still be
responsible for a significant fraction, as explained in the next
section. In certain models, these primordial
black holes (PBHs) are predicted to be produced at an interesting
level in early phases of the Universe due to density fluctuations
of the type also observed in the cosmic microwave background.
Their gravitational radius $r_g$ would be subatomic, $r_{g}\sim
10^{-8}\cdot \left(m_{pbh}/(10^{20}\ \rm{g})\right)$~cm. The
Bondi-Hoyle accretion radius, on the other hand, can be
significantly larger.\\

In this paper, we check whether hypothetical PBH collisions with
Galactic objects may be observable. A detection of observational
signatures of such collisions would support the hypothesis that
PBHs may represent a significant fraction of the DM in the Galaxy.
We find that the PBH collisions discussed by~\cite{zhilyaev2007}
would produce no presently-observable signatures. To prove our
case, we show that even with the most optimistic conditions
assumed, hypothetical collisions of PBHs with Galactic objects
place no observational constraints in the PBH mass range $\sim
10^{16}$~g~$<m_{pbh}$~$<10^{26}$~g considered in this paper.

The assumptions made are as follows:

\begin{itemize}

\item The mass distribution of PBHs is a delta-function, i.e.\ all
  PBHs have the same mass. The mass is treated as
  a free parameter, and the whole range $\sim 10^{16}$~g~$<m_{pbh}$~$<10^{26}$~g is considered. 
Any other distribution would significantly decrease
  the event rate: if we assume that all PBHs have the same mass and
  then consider this mass as a free parameter, we maximize the event
  rate for each single mass value, as we assume that 100\% of the
  total mass is focused at one arbitrary value. For wider mass distributions, the
  total mass of the PBHs is less than 100\% for a given mass value,
  which always
  reduces the event rate. 

\item The entire energy from a single interaction is taken to be in the energy
  band of the potential detector. In reality, we expect the signal to be
  spread from radio wavelengths up to soft gamma-rays. Using the entire
  radiated energy therefore gives an upper limit for detectability.

\item A single photon is sufficient for detecting the event. Even for a
  background-free source, three photons are required for giving 90\%
  confidence significance (see \citep{pdb2008}) and so with a more
  realistic estimate, the detection probability would be further reduced
  by a factor of between 3 and more than ten.

\item The signal is radiated at full opening angle, i.e.\ $2\pi$.
A more realistic estimate would
  include consideration of the shock opening angle, reducing the expected
  observation probability.

\item We assume that the signal is always
strong enough that it can reach us over the entire thickness of the
Galactic disk, $h\approx300$~pc.

\end{itemize}

Further, we make a few assumptions that
greatly simplifies calculations but could only slightly (indeed
 insignificantly) influence the final result:

\begin{itemize}

\item Luminous matter is taken to be distributed uniformly in the Galactic
  disk and dark matter is taken to have constant density in the spherical
  halo. We will show that the enhanced density in the Galactic Center
  yields slightly better results, but with the same conclusion that
  detection of the events can be neglected.

\item We use a delta-function for the distribution of relative velocity
  between the luminous and dark matter, $v_0=2.2\cdot
  10^{7}$~cm/s. We will show in this paper that the reduced
  velocity at the central part of the Galaxy does not influence our result.
  As for the possibility of a capture at low velocities, we
  show in Section \ref{added-section} that this does not occur within the
  energy range being considered.

\item The PBH is taken to be uncharged.

\end{itemize}

For alternative approaches to explaining dark matter by extending the
standard model of particle physics to include new particles, see
e.g.~\cite{haber_kane1984,bertone2005,biermann_kusenko2006,hooper_profumo2007}.

In order to investigate the interaction of PBHs with objects in the Galaxy
and possible observational consequences, we need to ask the following
questions:

\begin{enumerate}
\item {\it How frequent are the events?}
\item {\it How energetic is each single event?}
\item {\it What are the possible signatures?}
\end{enumerate}

In the following sections, we answer each question separately for the five
most important object classes in the Galaxy, i.e.\ main sequence stars,
red giants (in particular their dense cores), white dwarfs and neutron
stars. The main intrinsic properties of those objects are listed in
table~\ref{object_properties} together with their abundances in the
Galaxy. Both the interaction probability and the energy loss strongly depend on the properties listed. Also, we consider collisions with the
Earth, in order to investigate the claim that the Tunguska event
might have been caused in this way (see \cite{jackson_ryan1973}).
\placetable{object_properties}

This paper is organized as follows. Section~\ref{pbh_constraints}
summarizes the already-existing
observational constraints. In Section~\ref{estimation}, interaction
probabilities in the Galaxy and energy loss of a single event are derived.
Possible signatures and the detectability of
such events are investigated in Section~\ref{signatures}. Section
\ref{added-section} discusses the possibility of capturing the PBH in a
stellar object. Finally, Section~\ref{conclusion} summarizes the
conclusions to be drawn from our calculations and discusses further
possibilities for detection requiring additional assumptions.
\section{Primordial black holes - constraints \label{pbh_constraints}}
Ever since the hypothesis of primordial black holes was first
established~\citep{hawking1971}, different theoretical and
observational constraints on their actual abundance in the
Universe have been discussed.  While theoretical constraints are
rather speculative and model dependent (see
e.g.~\citet{carr2005,khlopov2008}), the observational ones are far
more robust. They firmly constrain the abundance of primordial
black holes in two mass ranges. In this section, we briefly review
them. Since we are interested in PBHs as a possible contributor to
the dark matter, we only consider PBHs that can have survived
until today, that is those with initial mass above $5\cdot
10^{14}$~g.

\begin{figure}
\centering{
\epsfig{file=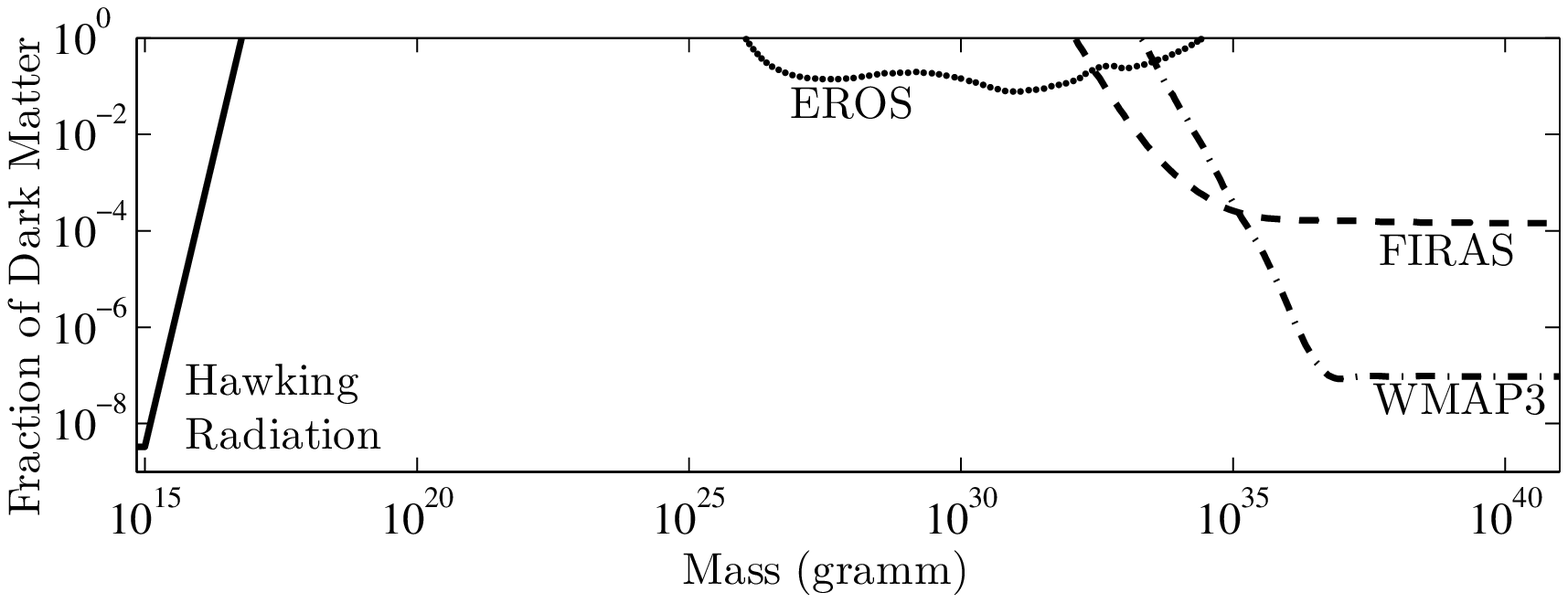,width=\linewidth}
\caption{Limits on the contribution to dark matter by PBHs. An absolute lower limit at around $10^{15}$~g is given by the fact that PBHs with lower
masses have evaporated by now. The limit at $\sim
  10^{15}$~g is derived from the background of photons at $>100$~MeV in the
  Universe, see~\citep{egret,barrau2003}.
  Micro-lensing constraints in the mass range
  $10^{26}$~g$<m_{pbh}<10^{33}$~g, come from the EROS
  experiment~\citep{alcock1998}. Limits in the
  mass range  $10^{33}$~g$<m_{pbh}<10^{40}$~g, derived from measurements of
  the Cosmic Microwave Background by FIRAS and WMAP, are presented in
  \cite{ricotti2008}. In this figure, we only show the most stringent
  limits in the different mass regions and disregard those limits with
  weaker constraints.
\label{constraints}}}
\end{figure}

 There are three main methods of looking for indirect detection of
primordial black holes above $5\cdot 10^{14}$~g. The first one aims for
the detection of PBH evaporation. This method is sensitive to the lower
mass limit of the initial mass function (IMF). This
initial mass function is expressed in
terms of the number of PBHs at their initial mass $m$, as a function of the
initial mass itself, $dn/dm$. As predicted
by~\citet{hawking1974}, and shortly afterwards also
by~\cite{bekenstein1975}, black holes evaporate due to quantum fluctuations
around the Schwarzschild radius of the black hole, leading to the
evaporation of PBHs with masses below $5\cdot 10^{14}$~g before the
present time. Therefore, non-detection of the high-energy photons which
would be produced by black hole evaporation, i.e.\ $E_{\gamma}\gtrsim
20$~MeV, leads to a robust constraint on the mass function around
$10^{15}$~g. At $10^{15}$~g, EGRET observations between $30$~MeV and
$120$~MeV imply that the PBH mass density must be smaller than $3.3\cdot
10^{-9}$ times that of the total dark
matter~\citep{egret,barrau2003}. Between $10^{15}-10^{16}$~g,
constraints from EGRET measurements are
still at a level of $<10^{-2}$ and provide significant constraints as well.
At significantly higher mass scales,
micro-lensing can help to restrict the abundance of PBHs, see
e.g.~\cite{afonso2003}. \cite{alcock1998} looked for planetary mass dark
matter via gravitational micro-lensing in the direction of the Large
Magellanic Cloud, resulting in constraints of up to 10\% in the mass range
$10^{26}$~g~$<m_{pbh}<10^{33}$~g. At even higher mass scales,
$10^{33}$~g~$<m_{pbh}<10^{40}$~g, results from anisotropy and spectra
measurements of the Comic Microwave Background (WMAP-3years and
COBE-FIRAS) can be used to constrain the number of PBHs
\citep{ricotti2008}. The excluded regions for the mass function are shown
in Fig.~\ref{constraints}. In addition, \cite{gould1992} proposed using
gamma ray bursts (GRBs) to detect ``femto-lenses'' produced by dark matter
objects in the mass range of $10^{17}$~g~$<m_{pbh}<10^{20}$~g. This method
has not yet been applied, however, and so no experimental limit has yet
been set. A mass range of about ten orders of magnitude,
$10^{15}$~g~$<m_{pbh}<10^{26}$~g, then remains at present unexplored.

In the standard scenario of density perturbations in the early Universe,
the IMF of PBHs, $dn/dm(m)$, behaves as, e.g.~\cite{halzen1991}
 \begin{equation}
\frac{dn}{dm}\propto m^{-2.5}\, ,
\end{equation}
 and so most PBHs have masses at around the current evaporation level. The
constraint by EGRET at the lower end of the mass scale therefore has
significant consequences for the abundance of black holes at higher
masses.\\

The above model for PBH production is not, however, the only one.
Several other proposed theories for PBH production predict significant
contributions from PBHs at different mass scales. Examples are the
collapse of topological defects, softening of the equation of state and
quantum gravity scenarios, which, although speculative, deserve to be
tested (see reviews by \cite{carr2005,khlopov2008} and references
therein).
 \placefigure{constraints}

\section{A toy model to estimates of the collision probability and
energetics\label{estimation}}
In order to describe possible observational signatures of the
hypothetical collisions of PBHs with stars in our Galaxy, one must
answer the two fundamental questions: {\it how frequent are the
collisions?} and {\it how energetic is a single collision?}
Precise answers to these questions are obviously very difficult to
obtain, as they depend on genuine astrophysical uncertainties that
cannot be resolved today. In the view of these uncertainties, we
estimate the collision frequency and energetics using a very
general and rather simple analytic toy model that adopts several
simplifying assumptions. For example, we assume that all PBHs have
the same mass $m_{pbh}$. We do not specify that value. Instead, we
consider $m_{pbh}$ to be one of a few free parameters of the
model.

Adopting ideas from the kinetic theory of gases, one may argue
that a reasonable estimate of the collision frequency between
stars and PBHs should depend only on the size $r$ of the region of
interest (e.g. either the whole Galaxy, or its central bulge),
numbers of stars $N_*$ and PBHs $N_{pbh}$ in this region, the
relative velocity $v_\star$ between stars and PBHs, and the collision
cross section $\sigma_{pbh,*}$. Similarly, a reasonable estimate
of the collision energetics should depend only on the mass and
radius of the star, $M_{\star}$, $R_{\star}$, mass of the black hole
$m_{pbh}$, and the relative velocity $v_\star$.

The final results obtained this way fully justify the adopted
strategy. They show that the hypothetical collisions would be far
too infrequent, or far too energetically weak, to be detected.
This holds {\it in the whole range} of $m_{pbh}$ and all other
parameters of the model. The mismatch involves several orders of
magnitude. Thus, our conclusion that {\it there are no
observational constraints from hypothetical collisions of
hypothetical dark halo primordial black holes with Galactic
objects} cannot be changed by including fine details, for example
by considering some particular mass functions for $m_{pbh}$.
\subsection{How frequent are the collisions?\label{eventrate}}
Let us consider a spherical part of the Galaxy with the radius $r$
containing $N_{\star}(r)$ stars. We will consider stars of a specific
type, e.g. main sequence, white dwarf, or neutron star. The dark
matter mass in the region is $M_{DM}(r)$. The number density of
primordial black holes can be estimated by assuming that PBHs
represent a fraction $\eta$ of the dark matter in the region, and
that the mass of a single PBH is $m_{pbh}$ ($\eta$ and $m_{pbh}$
are free parameters of the model),
 \begin{equation}
n_{pbh}(r)=\frac{\eta\cdot M_{DM}(r)}{4\,\pi/3\,r^{3}}\cdot
\frac{1}{m_{pbh}}\,.
\end{equation}
The collision frequency is given by,
 \begin{equation}
\dot{n}(r)\approx N_{\star}(r)\cdot j_{pbh}(r)\cdot \sigma_{pbh,\star}(r)
\approx N_{\star}(r)\cdot \frac{\eta\cdot
M_{DM}(r)}{4/3\,\pi\,r^{3}}\cdot \frac{V(r)}{m_{pbh}}\cdot
\sigma_{pbh,\star}(r)\,. \label{ndot:equ1}
\end{equation}
Here $\sigma_{pbh,\star}$ is the cross section of the interaction,
and $j_{pbh}(r) = n_{pbh}(r)\cdot V(r)$ is the flux of PBHs that move
with respect to stars with the relative velocity $V(r)$. We
approximate $V(r)$ by the rotational velocity in the Galaxy, which is
close to radial.

The cross section $\sigma_{pbh,\star}(r) = \pi\,R_{0}^{2}(r)$ should be
calculated from the energy and momentum conservation during the
encounter of gravitationally interacting star and PBH. The impact
parameter $R_{0}(r)$ of the encounter is defined by the PBH angular
momentum, equal to
\begin{equation}
m_{pbh}\cdot V(r)\cdot R_{0}(r)\,. \label{ang-mom1}
\end{equation}
As we are interested in an optimistic estimate of the event rate, and
hence of the cross section, here we use the conservation of angular
momentum in order to put an upper limit on the impact parameter $R_0$
for which there is a collision. This means that we use the maximum
value for $R_0$ for each collision, even if this value may be smaller
in reality. This maximum value for the impact
parameter corresponds to the limiting case of a PBH only grazing the
star's surface, so that from angular momentum conservation we get
\begin{equation}
 m_{pbh} V(r) R_0=m_{pbh} v_{\star}  R_{\star} \Rightarrow
 R_0=R_{\star} \frac{v_*}{V(r)}
\label{ang_mom}
\end{equation}
where $v_{\star}$ is the final PBH speed, that can be computed by considering
energy conservation in the limit of $m_{pbh} \ll M_{\star}$:
\begin{equation}
\frac{1}{2}V^2-\frac{G M_{\star}}{< R >}=\frac{1}{2}v_{\star}^2-\frac{GM_{\star}}{R_{\star}}
\end{equation}
with $M_{\star}$ being the mass of the star and $< R >$ being the typical distance between stars.
The characteristic distance at
which the PBH enters the gravitational field of the star can be
approximated as the typical distance between objects in the
Galaxy, assuming that each star-like object deflects the PBH at
least marginally. We take this typical distance to be $<R>\sim
1$~pc. We find that the gravitational energy is negligible in the
initial state\footnote{For $<R>=1$~pc, we get
  $G\,M_{\star}/<R>\approx 7\cdot 10^{3}$~cm\,s$^{-1}$.
    The initial gravitational potential only becomes important for
    distances below $<R> < 0.01$~pc, where the term $G\,M_{\star}/<R>$
  becomes comparable to the initial velocity. Since such short distances
  between stars are rather rare, we neglect this effect. As a
  comparison, the distance between the Sun and its closest neighbor is
  $\sim 1.3$~pc.}, so that the final velocity can be expressed as
\begin{equation}
v_{\star}\approx
\sqrt{V^{2}(r)+\frac{2\,G\,M_{\star}}{R_{\star}}}=\sqrt{V^{2}(r)+v_{esc}^{2}}\,,
\label{vstar}
\end{equation}
where $v_{esc}=2\,G\,M_{\star}/R_{\star}$ is the escape velocity
of the star. Combining equations (\ref{vstar}) and (\ref{ang_mom})
one arrives at
\begin{equation}
R_{0}(r)=R_{\star}\cdot \frac{v_{*}(r)}{V(r)}=R_{\star}\cdot
\frac{\sqrt{V^{2}(r)+v_{esc}^{2}}}{V(r)}\,.
\label{impact_parameter}
\end{equation}

Replacing the impact parameter
$R_0$ in Equ.\ (\ref{ndot:equ1}) with the help of Equ.\ (\ref{impact_parameter}) we get, finally
\begin{equation}
\dot{n}(r)\approx N_{\star}(r)\cdot \frac{\eta\cdot
M_{DM}(r)}{4/3\,r^{3}}\cdot \frac{V(r)}{m_{pbh}}\cdot
\left(\frac{v_{\star}(r)}{V(r)}\right)^{2}\cdot R_{\star}^{2}
\label{ndot:equ2}
\end{equation}
Now, the radial dependence of the parameters needs to be specified.
The observed velocity profile of the Galaxy can roughly be
approximated as $V(r)=v_0\cdot r/r_D$ for $0<r<r_d\approx 2.5$~kpc,
and $V(r)=v_0$ for $r_d\leq r<r_{halo}\approx 50$~kpc. Starting from
this observational fact, mass densities and masses enclosed in a
radius $r$ can be determined, as listed in table
\ref{table-regions}. This naturally divides our estimates in a central
part, considering only the bulge of the Galaxy, and an outer part,
considering the part of the Galaxy where the velocity is
constant. Here, we will first consider the outer part of the
Galaxy. In Section \ref{scaling-bulge}, we will show how our
results change for the central part.

\placetable{table-regions}

If the region considered is the whole outer part of the Galaxy
(the ``flat rotational curve region'' outside the bulge), one may
use $r = r_{halo} \sim 50$~kpc, $M_{DM}\approx 9\cdot
10^{11}\,M_{\odot}$, and $V(r)=v_0\approx 2.2\cdot
10^{7}$~cm\,s$^{-1}$. Introducing units convenient in this region,
$N_{\star,9}\equiv N_{\star}/10^{9}$,
$v_{\star,0}\equiv v_{\star}/v_0$, $R_{\star,9} \equiv
R_{\star}/(10^{9}{\rm cm})$ and $m_{pbh,20}\equiv
m_{pbh}/(10^{20}\,{\rm g})$, one writes (\ref{ndot:equ2}) in the
form
\begin{equation}
\dot{n}\approx 2\cdot 10^{-3}{\rm yr}^{-1}\cdot N_{\star,9}\cdot
\eta \cdot v_{\star,0}^2\cdot R_{\star,9}^{2}\cdot \left(\frac{v_0}{2.2\cdot
10^{7} {\rm cm/s}}\right)\cdot
m_{pbh,20}^{-1}\,.
\end{equation}
Apart from the PBH mass (which is a free parameter) this estimate
depends only on the stellar properties. For instance, white dwarfs
(with $v_{{\star}}/v_0=51/2.2$) have an event rate of
$1.4$~yr$^{-1}$ for a black hole mass of $10^{20}$~g.
\subsection{How energetic is a single event?\label{energy_deposit}}

In estimating the kinetic energy which is transferred from the black hole
to the star and then radiated, we follow previous studies of the motion of
accretors in a continuous medium, see e.g.~\citet{ruderman_spiegel1971}.
In making this first estimate, we assume a uniform and frictionless medium
and disregard self-gravity of the gas.

The black hole passing through the gas focuses matter behind itself,
creating a wake, because of the gravitational interaction transferring
momentum to the atoms nearby. The inhomogeneity in the medium gives rise
to a force acting on the black hole, causing a decrease in kinetic energy.

As pointed out by~\citet{ostriker1999}, the force on the black hole (and
hence the energy loss) depends on whether the motion is supersonic or not,
as well as on the degree of supersonic motion. In making a first order
approximation, we consider the black hole to be moving at constant speed.
Also, we assume the velocity to be supersonic, so that the drag force, and
hence the energy loss, reduces to that given
by~\citet{ruderman_spiegel1971}. The energy loss due to the bow shock is
negligible compared to that due to the tail shock, and the energy loss per
time is then given by
 \begin{equation}
\frac{dE_{\rm loss}}{dt}=
\frac{4 \pi G^2 m_{pbh}^{2} \rho_{\star}}{v_{{\star}}}
\ln{\frac{D_{\max}}{D_{\min}}}\,,
\label{eq:power_loss}
\end{equation}
 where $D_{\max}$ is the maximum linear dimension of the star (i.e.~its
diameter) and $D_{\min}$ is the linear dimension of the accretor, lying
between the black hole's horizon radius and its Bondi-Hoyle accretion radius. The
exact formulation of the linear scales is not crucial for this first
estimate, as they appear in a logarithm. Here, we use
$\ln(D_{\max}/D_{\min})=10$. The total energy loss is then given by
 \begin{equation}
E_{\rm loss}=\frac{dE_{\rm loss}}{dt}\cdot dt=
\frac{8 \pi G^2 m_{pbh}^{2} R_{\star} \rho_{\star}}{v_{{\star}}^{2}}
\ln{\frac{D_{\max}}{D_{\min}}}\,,
\label{eq:en_loss}
\end{equation}
 considering a time interval of $dt\approx 2\cdot
R_{{\star}}/v_{{\star}}$. Equation~(\ref{eq:en_loss}) can be expressed as
 \begin{equation}
E_{\rm loss}=2\cdot 10^{27}\,{\rm erg}\,\cdot m_{pbh,20}^{2}\cdot
R_{\star,9}\cdot \left(\frac{v_{\star}}{2.2\cdot 10^{7}\,{\rm cm/s}}\right)^{-2}\cdot
\rho_{\star,5}\,,
\label{eq:en_loss_num}
\end{equation}
 where $\rho_{\star,5}\equiv\rho_{\star}/(10^{5}\,{\rm g\,cm^{-3}})$.
Table~\ref{energy:table} lists the energy released and the interaction
probability for black holes of $10^{20}$~g passing through the different
types of Galactic objects.
 \placetable{energy:table}

In our picture, the shock front will locally heat the plasma to an extent
depending on the particular Galactic object under consideration. The
spectrum is therefore a thermal one. In the following, we discuss the peak
energy of the spectrum in order to estimate the detectability for
detector systems observing in the relevant energy range.

To calculate the temperature of the shock, we assume supersonic shock
conditions and a perfect gas, giving
 \begin{equation}
T_S= \frac{5}{64} \frac{m_H}{k_B} {v_{\star}}^2 M^2
>4.5\,\cdot 10^5 \,{\rm K}\,\cdot  \left(\frac{v_{\star}}{2.2\cdot 10^{7}\,{\rm cm/s}}\right)^2 \,,
\end{equation}
 where the downstream velocity is diminished by a factor of four with
respect to the upstream velocity (see, for
example,~\citet{landau_lifshitz_fluid,parks1991}). Here, $k_B$ is the
Boltzmann constant, $v_{\star}$ is the velocity of the shock, $m_H$ is the
mass of a hydrogen atom, $M$ is the Mach number, and the average relative
velocity between primordial black holes and stars is taken to be $2.2\cdot
10^{7}$~cm~s$^{-1}$. Wien's law now gives the frequency at maximum
intensity as
 \begin{equation}
\nu_{\max}=\frac{c\,T_S}{b}=\frac{c}{b}\cdot\frac{5}{64}\frac{m_H}{k_B}
{v_{\star}}^2 M^2
>4.5\cdot 10^{16}\,{\rm Hz} \,\cdot  \left(\frac{v_{\star}}{2.2\cdot 10^{7}\,{\rm cm/s}}\right)^2  \,,
\end{equation}
 where $b=2.9 \cdot 10^{- 3}$~Km. This can also be expressed in terms of
the peak photon energy
 \begin{equation}
E_{\rm peak}=h \nu_{\max}
>  0.2\,{\rm keV}\,\cdot  \left(\frac{v_{\star}}{2.2\cdot 10^{7}\,{\rm cm/s}}\right)^2 \,.
\label{epeak_star}
\end{equation}
 This is merely a lower limit, but should suffice as a first-order
estimate. For a collision with Earth, an extreme ultraviolet spectrum is
expected, while collisions with main sequence stars and red giant cores
are expected to give an X-ray spectrum. White dwarfs and neutron stars
cannot be treated with simple thermodynamics: we here assume that the
shock velocity $v_{\star}$ determines the temperature of the shock.
Thus, we get a peak energy which is increased by a factor $4^2=16$,
\begin{equation}
E_{\rm peak}= 1.6\,{\rm keV}\,\cdot  \left(\frac{v_{\star}}{2.2\cdot 10^{7}\,{\rm cm/s}}\right)^2 \,.
\label{epeak_rel}
\end{equation}
For a white dwarf,
we get a soft $\gamma$ spectrum, while for a neutron star we get
high-energy gamma rays, see table~\ref{energy:table}. Since we are
dealing with order-of-magnitude estimates here, the actual peak energy
could deviate from what is calculated. Therefore, we present our
results for all objects and for all possibly interesting energy
ranges, i.g.\ X-rays up to $\gamma-$rays.
%
%
\subsection{A scaling to the bulge \label{scaling-bulge}}
%
%
The estimates of frequency (\ref{ndot:equ2}) and energy
(\ref{eq:en_loss}) are general and valid everywhere in the Galaxy.
We have already numerically specified them for the region outside
the bulge, where the velocity $v_0 = V(r) = {\rm const}$. We will
now discuss them for the central bulge. Note that only the
frequency equation (\ref{ndot:equ2}) contains parameters $r$, $N_*
\approx M_{lum}/M_{\odot}$, $M_{DM}$, $v_0$ that depend on the
region in the Galaxy. The energy equation (\ref{eq:en_loss}) does
not depend on the relative velocity of stars and PBHs $v_0$
because the actual velocity of collision $v_{\star}$ is usually
dominated by the
escape velocity at the stellar surface.

The frequency equation (\ref{ndot:equ2}) scales with the relevant
parameters as,
\begin{equation}
{\dot n} \sim M_{lum}\,M_{DM}\, r^{-3}\,v_0^{-1}\,.
\label{scaling-frequency}
\end{equation}
Let us now consider a model of the Galaxy in which the rotational
velocity $V(r)$ is proportional to the distance from the Galactic
Center in the ``bulge'' and constant ``outside''. This
approximately follows what is observed. Distributions of the total
(i.e. luminous plus dark matter) mass $M_{tot}(r)$ and its density
$\rho(r)$ may be calculated from
\begin{equation}
\frac{V^2}{r} = \frac{d\Phi}{dr}, ~~-4\,\pi G \rho =
\frac{1}{r^2}\frac{d}{dr}\left( r^2 \frac{d\Phi}{dr}\right),
~~\frac{dM_{tot}}{dr} = 4\pi r^2\,\rho.
\label{galactic-model}
\end{equation}
Results are given in Table \ref{table-regions}. One may assume
that $M_{lum}(r)\,M_{DM}(r) \sim M_{tot}^2(r)$. Then, in the bulge it is
$M_{tot}^2 \sim r^6$, $V(r) \sim r$ and therefore ${\dot n} \sim r^{2}$,
which means that in the bulge the frequency of collisions is
maximal for $r = r_D$; it does {\it not} increase towards the
bulge center. Let us denote this maximal frequency by ${\dot
n}_D$, and let us denote the frequency in the outside region by
${\dot n}_{out}$. From the above discussion and the last column of
Table \ref{table-regions} it follows that,
\begin{equation}
\frac{{\dot n}_D}{{\dot n}_{out}}
= \frac{M^2_D\, r^{-3}_D}{M^2_{out}\, r^{-3}_{out}}
= \frac{r_{out}}{r_{D}} \approx 20.
\label{frequency-ratio}
\end{equation}
In the view of our estimate at the end of Section \ref{eventrate},
the collision frequency ${\dot n}_{out}$ in the outside part of
the Galaxy is extremely small. The factor of $20$ does change much
the situation in the bulge; the collision frequency ${\dot n}_{D}$
is also very small there. For this reason, in the rest of this
paper we will not distinguish between the outside and the bulge,
and take in our estimates of observational detectability ${\dot n}
= {\dot n}_{out}$ as representative for the whole Galaxy.
\section{Signatures and detectability of a single encounter \label{signatures}}

An encounter in which a PBH passes through a stellar object would be
detectable if the following conditions are met:
 \begin{enumerate}
\item {\it each single event is strong enough to be seen by the detector;}
\item {\it the events are frequent enough so that they can be seen within
  the detector's life time.}
\end{enumerate}

In subsection \ref{dmax}, we use the sensitivity of current detectors
to calculate the maximum distance at which an event could be observed; in
subsection \ref{tobs:sec}, we estimate the required observation time for a
detection. Both estimates are based on the energy loss and event rate
calculated in Sections \ref{energy_deposit} and \ref{eventrate}. In
Section \ref{detectability:sec}, we determine the actual observation time
required in order to see a single event in one of the current detectors.
 \placetable{detectors:table}
\subsection{Detector sensitivity and photon fluence\label{dmax}}
The aim of this section is to calculate the maximum distance at
which an encounter involving a PBH passing through a Galactic object could
be observed by a detector. This can be done by comparing the sensitivity
of a detector, $F_{sens}$, to the photon fluence from the event,
$F_{\gamma}$. Detection is only possible for
 \begin{equation}
F_{\gamma}\geq F_{sens}\,.
\label{fgamma_fsens:1}
\end{equation}
 The photon fluence and instrumental sensitivity can be determined as follows:
\begin{itemize}
 \item {\bf Photon fluence from a PBH passing through a Galactic object}\\
 The photon fluence at Earth depends on the total energy deposited as
given by Equ.\ (\ref{eq:en_loss_num}) and on the distance of the event from
Earth, $d$,
 \begin{equation}
F_{\gamma}=\frac{E_{\rm loss}}{\Omega\cdot (1+z)\cdot
  d^{2}}\approx\frac{E_{\rm loss}}{\Omega\cdot d^{2}}\,.
\label{fgamma}
\end{equation}
 The redshift factor $(1+z)$ describes adiabatic energy losses, which can
be neglected on Galactic scales, $z\approx 0$. The opening angle of the
emitted signal $\Omega$ is determined by the shape of the shock.

\item {\bf Sensitivity}\\
 For a positive detection, an instrument with an effective area $A_{eff}$
needs to see at least one photon of energy $E_{\gamma}$, giving a fluence
sensitivity of
 \begin{equation}
F_{sens}=\frac{E_{\gamma}}{A_{eff}}\, .
\label{fsens}
\end{equation}
 This assumes that there is no additional background; as soon as there is
some significant background, more than one photon would be needed for
detection. We use this single photon argument to give an absolute lower
limit on the fluence required in order for the signal to be detected.
 \end{itemize}
 Combining Equations (\ref{fgamma_fsens:1}), (\ref{fgamma}) and
(\ref{fsens}), we have
 \begin{equation}
\frac{E_{\rm loss}}{\Omega\cdot d^{2}}\geq
\frac{E_{\gamma}}{A_{eff}}\,.
\end{equation}
 Typically, the effective detection area $A_{eff}$ depends on the energy
$E_{\gamma}$ of detection, and the detector is sensitive to the shape of
the spectrum. Here, we optimistically assume that the effective area is
concentrated at the lower energy threshold of the detector. For higher
energies, the sensitivity would be less. For a given detector, the
distance between the detector and the event to be observed must be less
than
 \begin{equation}
d \leq \sqrt{\frac{E_{\rm loss}}{E_{\gamma}}\cdot \frac{A_{eff}}{\Omega}}=:d_{\max}\,.
\label{eq:dmax}
\end{equation}
 Inserting into this the expression for the total energy released, as
given by Equ.~(\ref{eq:en_loss_num}), leads to
 \begin{eqnarray}
d \leq d_{\max}=4.7\,{\rm pc}&\cdot&m_{pbh,20}\cdot R_{\star,9}^{1/2}\cdot \left(\frac{v_{\star}}{2.2\cdot 10^{7}\,{\rm cm/s}}\right)^{-1}\cdot
\rho_{\star,5}^{1/2}\cdot A_{eff,2500}^{1/2}\cdot
\Omega_{2\pi}^{-1/2}\cdot
E_{\gamma,1.5}^{-1/2}\,,
\label{dmax:equ}
\end{eqnarray}
 where values are given in terms of the X-ray detector XMM Newton: the
effective area reference value is $A_{eff,2500}\equiv A_{eff}/(2500\,{\rm
cm}^2)$, at a reference energy of $E_{\gamma,1.5}\equiv E_{\gamma}/(1.5\,{\rm
keV})$. The opening angle of the signal is expressed in terms of the
maximum opening angle, corresponding to a signal emitted over half a hemisphere,
$\Omega_{2\pi}\equiv \Omega/(2\,\pi)$. 

\subsection{Observation time and interaction probability \label{tobs:sec}}
 We can evaluate the observation time $t_{obs}$ required for detecting the
effect of a PBH passing through a Galactic object, by considering the
event rate within the observable distance $d_{\max}$, $n(d_{\max})$. The
latter can be calculated from the total event rate, $\dot{n}$, as
calculated in Section \ref{eventrate}:
 \begin{equation}
\dot{n}(d_{\max})=\frac{V(d_{\max})}{V_{tot}}\cdot \dot{n}\,,
\label{ndmax}
\end{equation}
 where, $V(d_{\max})$ is the volume containing the observable events
\begin{equation}
V(d_{\max})=\pi\,d_{\max}^2\,h
\label{vdmax}
\end{equation}
and
\begin{equation}
V_{tot}=\pi\,d_{\rm Galaxy}^{2}\,h\,,
\label{vtot}
\end{equation}
 with $d_{\rm Galaxy}\approx 15$~kpc being the radius of the Galaxy. We
assume an isotropic distribution of sources within the Galactic disk with
a height $h=300$~pc. This means that we assume that the signal is always strong
enough to reach Earth from $300$~pc across the height of the disk. Realistically, some signals will
be lost, since they are too weak to be detected at Earth, but as mentioned before, we work with the
most optimistic assumptions in this paper.

The number of events occurring within the observable distance $d_{\max}$
from Earth is given as:
 \begin{equation}
N=\dot{n}(d_{\max})\cdot \frac{\Omega_{obs}}{4\pi}\cdot
\frac{\Omega}{4\pi}\cdot t_{obs}\,.
\label{N}
\end{equation}
 In this equation, the solid angle $\Omega_{obs}$ describes the Field of
View (FoV) of the detector. The solid angle $\Omega$ accounts for a
possibly beamed signal, which reduces the total number of events to be
observed, since only a fraction $\Omega/(4\pi)$ is directed towards Earth.
The total event rate $\dot{n}$ in the Galaxy was determined in
Section~\ref{eventrate}.

Using Equations (\ref{ndmax}), (\ref{vdmax}) and (\ref{vtot}) in Equ.\
(\ref{N}), the number of events within a radius $d_{\max}\leq d_{\rm
Galaxy}$ is given as
 \begin{equation}
N=\dot{n}\cdot \left(\frac{\min\left(d_{\max},d_{\rm Galaxy}\right)}{d_{\rm Galaxy}}\right)^{2}\cdot
\frac{\Omega_{obs}}{4\pi}\cdot \frac{\Omega}{4\pi}\cdot t_{obs}\,.
\end{equation}
 The radius of the Galaxy $d_{\rm Galaxy}$ gives an absolute limit for the
number of events. For detection of a single event ($N=1$), the required
observation time is then
 \begin{equation}
t_{obs}=1\,{\rm yr} \cdot \frac{4\pi}{\Omega_{obs}}\cdot \frac{4\pi}{\Omega}\cdot
\left(\frac{\dot{n}}{1\,{\rm yr}^{-1}}\right)^{-1}\cdot
\left(\frac{\min\left(d_{\max},d_{\rm Galaxy}\right)}{d_{\rm Galaxy}}\right)^{-2}\,.
\label{tobs}
\end{equation}
 Combining Equations (\ref{ndot:equ2}), (\ref{dmax:equ}) and (\ref{tobs})
gives
 \begin{eqnarray}
t_{obs}(m_{pbh}<m_{pbh}^{\rm break})=7.0\cdot 10^{11}\,{\rm
  yr}\,&\cdot&\, m_{pbh,20}^{-1}\cdot
R_{\star,9}^{-3}\cdot
\rho_{\star,5}^{-1}\cdot N_{\star,9}^{-1}\cdot \Omega_{2\pi}^{-1} \nonumber\\
&\cdot& \eta^{-1}\cdot A_{eff,2500}^{-1}
\cdot
\Omega_{obs,0.2}^{-1}\cdot
E_{\gamma,1.5}
\end{eqnarray}
 for $d_{\max}<d_{\rm Galaxy}$ and
\begin{eqnarray}
t_{obs}(m_{pbh}\geq m_{pbh}^{\rm break})=1\cdot 10^{3} \,{\rm yr} &\cdot& \Omega_{obs,0.2}^{-1}\cdot
\Omega_{2\pi}^{-1}
\cdot N_{\star,9}^{-1}\cdot
\eta^{-1}\nonumber\\
&\cdot &m_{pbh,20}\cdot v_{\star,0}^{-2}\cdot R_{\star,9}^{-2}\cdot
\left(\frac{v_0}{2.2\cdot 10^{7}\,{\rm cm/s}}\right)^{-1}
\end{eqnarray}
 for $d_{\max}=d_{\rm Galaxy}$. Here, $\Omega_{obs,0.2}\equiv
\Omega_{obs}/(0.2\,{\rm sr})$.

This break in the function $t_{obs}(m_{pbh})$ occurs at a mass of
\begin{equation}
m_{pbh}^{\rm break}=3.2\cdot 10^{23}\,{\rm g}\,\cdot\,
R_{\star,9}^{-1/2}\cdot \left(\frac{v_{\star}}{2.2\cdot 10^{7}\,{\rm cm/s}}\right)\cdot\rho_{\star,5}^{-1/2}
\cdot
\Omega_{2\pi}^{1/2}
\cdot A_{eff,2500}^{-1/2}\cdot E_{\gamma,1.5}^{1/2}\,.
\end{equation}

The observation time reaches its minimum at this break mass. At
  lower masses up to the break mass, $m_{pbh}\leq m_{pbh}^{\rm break}$,
  the observation time follows a $m_{pbh}^{-1}-$ behavior. At higher
  masses, $m_{pbh}>m_{pbh}^{\rm break}$, the observation time increases as
  $m_{pbh}^{2}$. Observation conditions are therefore optimal at the break
  mass.
\subsection{Detectability\label{detectability:sec}}
 As discussed in Section~\ref{energy_deposit}, the signals are expected to
come at X-ray energies. We focus here on four detectors currently in
operation which are well-suited for these investigations because of their
relatively large FoVs and effective areas. First we consider {\sc XMM
Newton} which is sensitive from extreme UV to X-rays. We use its effective
area as quoted in~\cite{xmm_performance2002} at the reference energy
$1.5$~keV. The results which we obtain for this are equivalent to what is
expected for observations with Chandra. For higher energies, from hard
X-rays to soft gamma-rays, we use {\sc Swift-BAT} and {\sc Fermi-GBM} as
reference detectors. Swift covers the range between $15$~keV and
$150$~keV; the high-energy detector of the GBM covers the energy range
from $150$~keV to $30$~MeV. Both instruments provide quite large effective
areas in combination with a large FoV, since they were designed to detect
GRBs. For high-energy radiation, we use Fermi-LAT parameters, covering the
range from $20$~MeV up to $300$~GeV. Table~\ref{detectors:table}
summarizes the properties of the four detectors. For Swift and Fermi, we
use the lower energy thresholds as the reference energies for the
effective areas quoted
in~\citep{swift_performance,fermi_gbm2008,fermi_lat2003}. This is the most
optimistic approach, since using higher energies would lead to reduced
sensitivities.
 \placetable{detectors:table}

\begin{figure}
\centering{
\epsfig{file=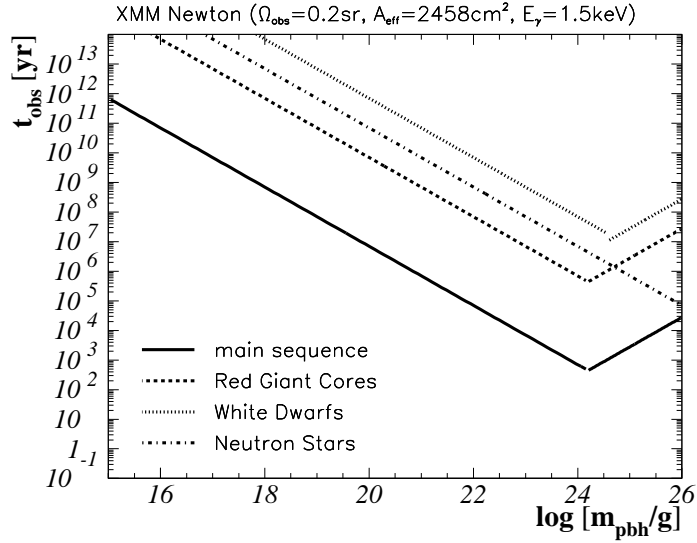,width=10cm}
\caption{Observation time of PBH interactions with Galactic objects,
  observing with XMM Newton. Solid line: main sequence stars; dashed line: red
  giant cores; dotted line: white dwarfs; dot-dashed line: neutron stars. The observation time decreases with
  $m_{pbh}^{-1}$, until the point where the entire Galaxy can be observed,
  beyond which it starts increasing with $m_{pbh}$.
  \label{tobs_incl_dmax_xmm:fig}}
}
\end{figure}

\begin{figure}
\centering{
\epsfig{file=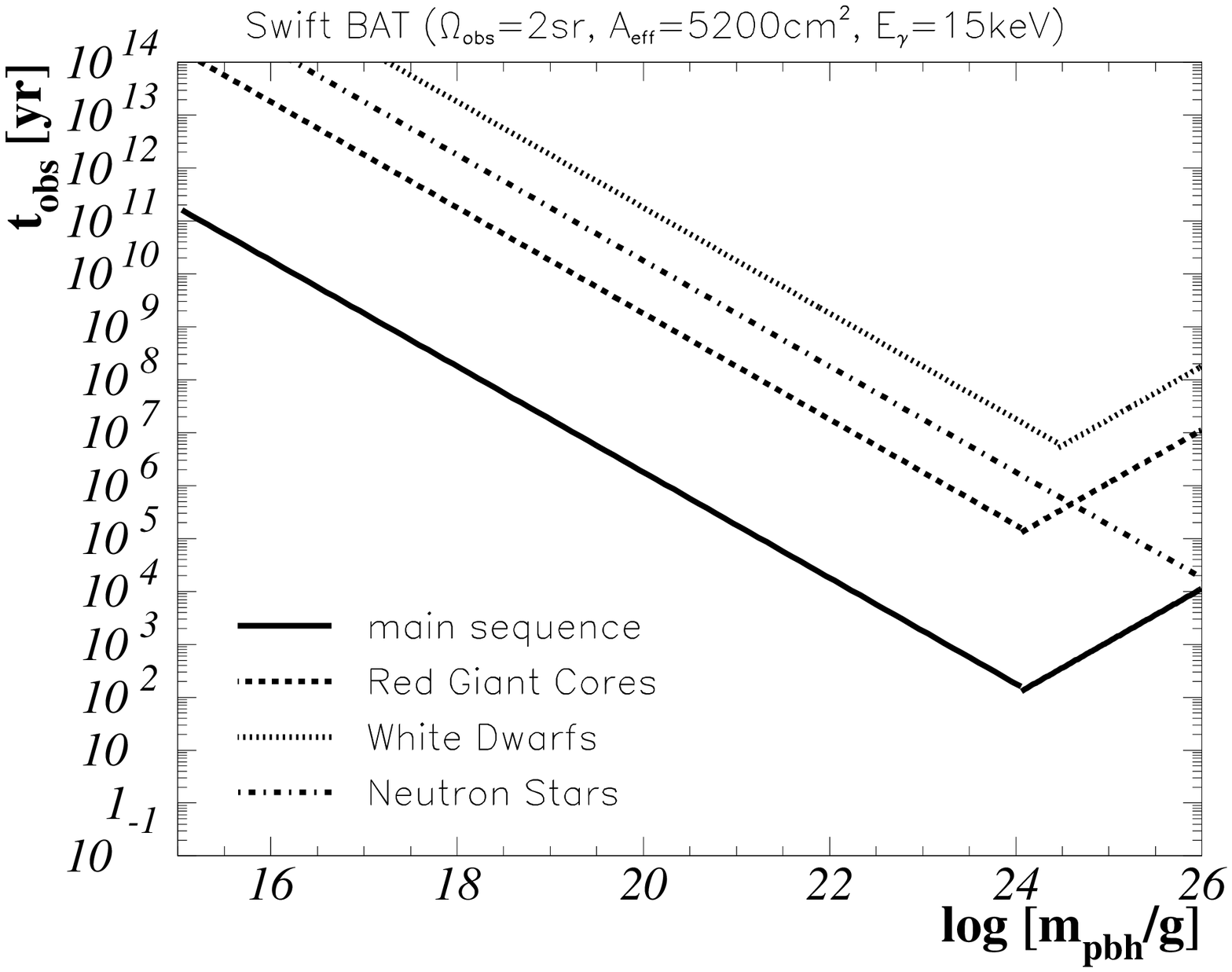,width=10cm}
\caption{ Observation time of PBH interactions with Galactic objects,
  observing with Swift-BAT. Same representation as in Fig.~\ref{tobs_incl_dmax_xmm:fig}.\label{tobs_incl_dmax_swift:fig}}
}
\end{figure}

\begin{figure}
\centering{
\epsfig{file=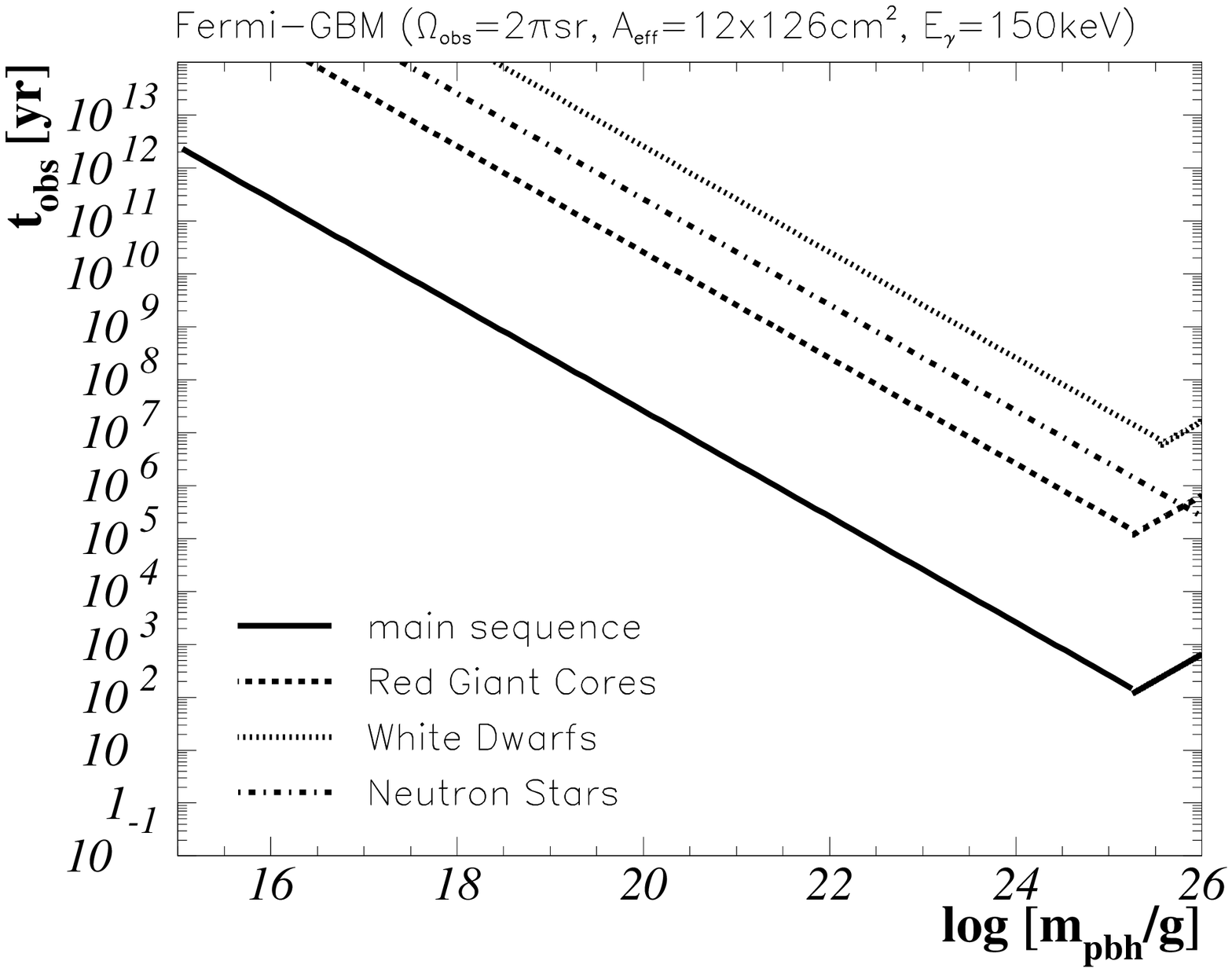,width=10cm}
\caption{Observation time of PBH interactions with Galactic objects,
  observing with Fermi-GBM. Same representation as in Fig.~\ref{tobs_incl_dmax_xmm:fig}.\label{tobs_incl_dmax_fermi_gbm:fig}}
}
\end{figure}

\begin{figure}
\centering{
\epsfig{file=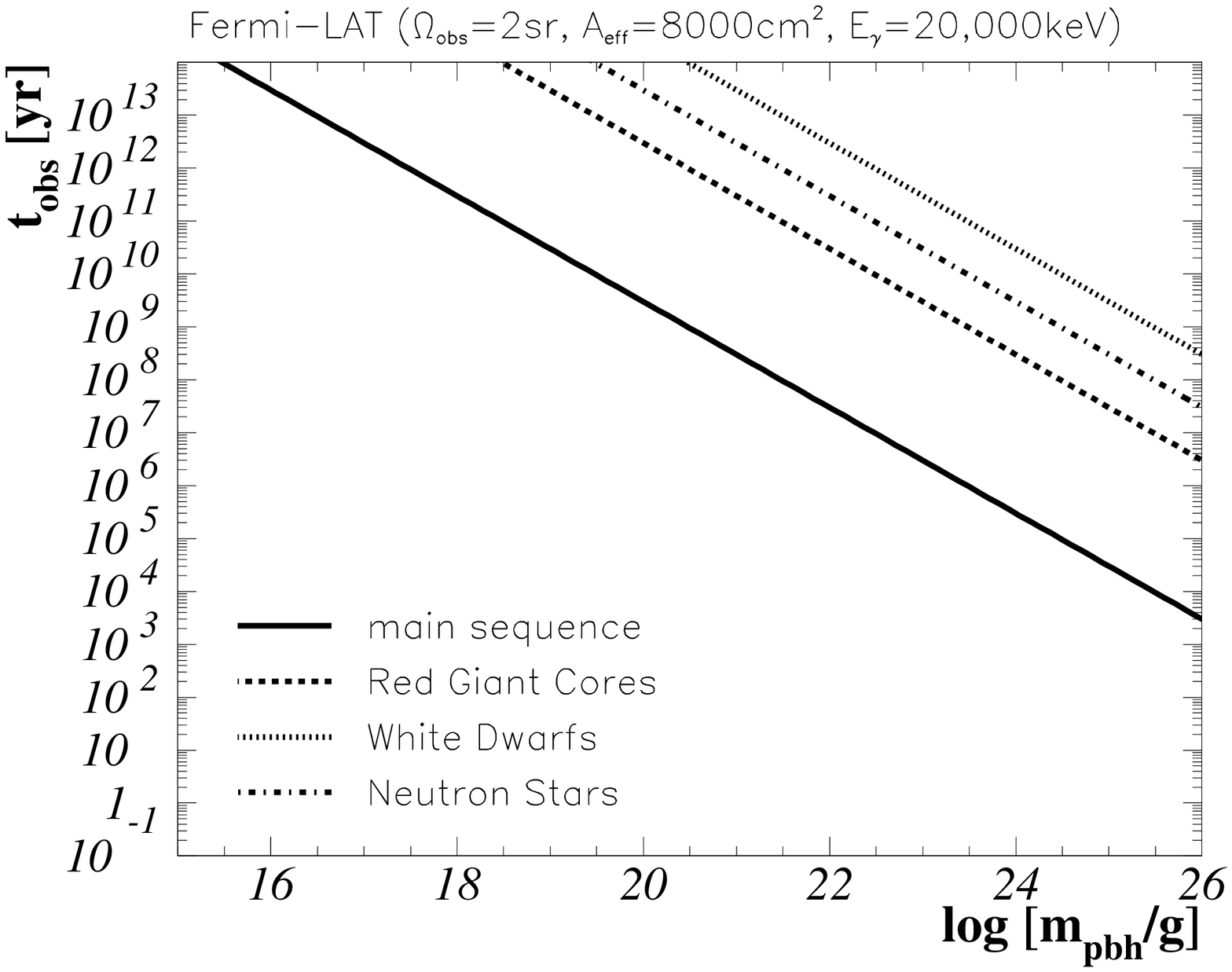,width=10cm}
\caption{Observation time of PBH interactions with Galactic objects,
  observing with Fermi-LAT. Same representation as in Fig.~\ref{tobs_incl_dmax_xmm:fig}.\label{tobs_incl_dmax_fermi_lat:fig}}
}
\end{figure}

The detector's effective area, FoV and reference energy then determine the
observation time and the break masses as discussed in the previous
subsection. It turns out that the observation times for the four different
detectors are related as
 \begin{equation}
t_{obs}^{\rm XMM}=2\cdot t_{obs}^{\rm Swift/BAT}=0.2\cdot t_{obs}^{\rm Fermi/GBM}=0.003\cdot t_{obs}^{\rm Fermi/LAT}
\end{equation}
 and the break masses $m_{pbh}^{\rm break}$ are related as
\begin{equation}
m_{pbh}^{{\rm break},\, \rm{XMM}}=2\cdot m_{pbh}^{{\rm break},\, {\rm
    Swift/BAT}}=0.1\cdot m_{pbh}^{{\rm break},\,{\rm Fermi/GBM}}=0.02\cdot
m_{pbh}^{{\rm break},\,{\rm Fermi/LAT}}\,.
\end{equation}
 Swift/BAT gives the shortest observation time and is therefore best
option concerning this. Fermi/LAT gives the largest break mass.
Figure~\ref{tobs_incl_dmax_xmm:fig} shows observation time plotted against
PBH mass for the case of XMM Newton in the relevant mass range of
$10^{15}$~g~$<m_{pbh}<10^{26}$~g. Masses outside this range are already
known not to contribute significantly ($<10\%$) to the dark matter. The
solid line represents main sequence stars, the dashed line shows red giant
cores, the dotted line is the result for white dwarfs and the dot-dashed
line is for neutron stars.
 \placefigure{tobs_incl_dmax_xmm:fig}
 The shortest observation times are obtained for main sequence stars, and
even there, the best case at $m_{pbh}=2\cdot 10^{24}$~g requires an
observation time of more than $100$~years. Similar results are obtained
for Swift, Fermi-GBM and Fermi-LAT, as shown in
Figures~\ref{tobs_incl_dmax_swift:fig}, \ref{tobs_incl_dmax_fermi_gbm:fig}
and \ref{tobs_incl_dmax_fermi_lat:fig}: here as well, the observation
times would need to be longer than $100$~years in order to see one event,
except in the case of Fermi/GBM at $3\cdot 10^{25}$~g, where $60$~years
would be sufficient. The best observation times $t_{obs}^{best}$ are those
for the break points in the plots of the observation time against PBH mass
$m_{pbh}^{\rm break}$. They are listed for the four detectors in
Table~\ref{detectors:table}.
 \placefigure{tobs_incl_dmax_swift:fig}
\placefigure{tobs_incl_dmax_fermi_gbm:fig}
\placefigure{tobs_incl_dmax_fermi_lat:fig}

These observation times are far longer than the lifetimes of the present
detectors and so, given that they are anyway optimistic lower limits on
the real times, we conclude that we cannot obtain meaningful constraints
in this way. The event rates are too low and the energy released is too
small.\\
Any constraint which we could give for the PBH contribution to
dark matter in the Galaxy would be above 100\%. We therefore do
not convert the observation times into limits on the dark matter
contribution.

\subsection{Alternative signatures}
 There are a few other processes that could in principle lead to the
production of a detectable signal for different detectors. We will
briefly discuss here why these mechanisms also fail to produce a
significant signal:
\begin{itemize}
\item {\it Shock acceleration of charged particles}\\
 Particles can gain energy by scattering off magnetic inhomogeneities via
the mechanism of stochastic acceleration~\citep{fermi1949,fermi1954}.
However, inside stars the densities are typically very high, and for white
dwarfs and neutron stars the matter is degenerate. The mean free path of a
proton in matter can be determined by considering the total cross section
for proton-proton interactions (with $\sigma_{p\,p}\approx 50$~mbarn):
 \begin{equation}
\lambda_{mfp}=\frac{m_p}{\rho_{\star}\cdot
  \sigma_{p\,p}}=3\cdot 10^{-4}\,{\rm cm}\,\cdot\, \rho_{\star,5}\,.
\end{equation}
 This mean free path is typically smaller than a centimeter and does not
allow for significant acceleration. Injection and acceleration of
energetic charged particles in shock fronts is therefore highly
disfavored, with the possible exception of the tenuous envelopes of red
giant stars.

\item {\it Star quakes}\\
  PBHs passing through an astrophysical object can induce star quakes.
  Although the emitted photons themselves cannot be distinguished from the
  general thermal spectrum of the star, the event could be detected by
  Fourier-transforming the spectrum from time to frequency space, see
  \cite[e.g.]{donea1999,martinez_oliveros2007} and references therein.
  However, these quakes can so far only be detected for the Sun and our
  rate for encounters between a PBH and the Sun is only $10^{-7}\,{\rm
  yr}^{-1}$. This method for getting detections can therefore be excluded.
\end{itemize}
\section{On the possibility of a PBH being captured by a star
\label{added-section}}
 \cite{turolla2009} have considered the possibility of a PBH being
captured by a star. In this section, we show that such a capture could
happen only for PBHs with masses greater than about $10^{28}$ g, i.e. in
the range already constrained by the micro-lensing results
discussed earlier. The capture rates would therefore be reduced by a factor of 10 or more,
considering that $\eta<0.1$.

The energy balance for a PBH being attracted by an astrophysical object
in the formulation of a reduced two-body problem is given by
 \begin{equation}
E_{\rm in}-E_{\rm loss}=E_{\rm out}\,.
\label{e_balance:equ}
\end{equation}
 Here, $E_{\rm loss}$ is the energy loss undergone by the PBH as it passes
through the star and $E_{\rm out}$ is its final energy. Its initial energy
is given by
 \begin{equation}
E_{\rm in}=\frac{1}{2}\cdot  m_{pbh} \cdot v_{0}^{2}\,,
\label{ein:equ}
\end{equation}
 where we replace the reduced two-body mass by $m_{pbh}$, because we are
considering \mbox{$M_{\star}>>m_{pbh}$}. The PBH can be captured by the
star only if a sufficient part of the its initial kinetic energy can be
dissipated (``lost'') by the effective dynamical friction as it passes
through the star. The condition for getting a bound system is $E_{\rm
out}<0$. From Equations\ (\ref{e_balance:equ}) and (\ref{ein:equ}) we get
 \begin{equation}
E_{\rm loss} > E_{\rm in}=\frac{1}{2}\cdot  m_{pbh} \cdot v_{0}^{2}\,.
\label{vel_ineq2:equ}
\end{equation}
 This is the condition that needs to be fulfilled in order to have a
capture. The energy loss from effective friction should be dominant, as
determined from Equ.\ (\ref{eq:en_loss}). It follows that
 \begin{equation}
\frac{2\,E_{\rm loss}}{m_{pbh}}=4\cdot
10^{7}\,\frac{\rm cm^2}{\rm s^2}\,\cdot m_{pbh,20}\cdot
\left(\frac{v_{\star}}{2.2\cdot 10^{7} {\rm cm/s}}\right)^{-2}\cdot R_{\star,9}\cdot
\rho_{\star,5}\stackrel{\rm (Equ.\ \ref{vel_ineq2:equ})}{>}v_{0}^{2}\,.
\end{equation}
 We can now derive a condition for the minimum PBH mass required for
capture:
 \begin{eqnarray}
m_{pbh,20}&>&\left(\frac{v_{0}}{2.2\cdot 10^{7} {\rm cm/s}}\right)^{2}\cdot \left(\frac{v_{esc}^{2}+v_{0}^{2}}{4\cdot 10^{7} {\rm
    cm^2/s^2}}\right)\cdot R_{\star,9}^{-1}\cdot \rho_{\star,5}^{-1}\\
&>&\left(\frac{v_{0}}{2.2\cdot 10^{7} {\rm cm/s}}\right)^{2}\cdot\left(6.6\cdot 10^{9}\cdot M_{\star,\odot}\cdot
  R_{\star,9}^{-1}+1.2\cdot 10^{7}\cdot\frac{v_{0}^{2}}{\left(2.2\cdot 10^{7} {\rm
    cm/s}\right)^2}\right)\cdot R_{\star,9}^{-1}\cdot \rho_{\star,5}^{-1}\,.
\end{eqnarray}
 The minimum mass for which capture can occur can now be computed for each
of the astrophysical objects which we are considering (see table
\ref{energy:table}):
 \begin{eqnarray}
m_{pbh}&>&
6.6\cdot 10^{28}\,{\rm g}\cdot\frac{v_{0}^{2}}{(2.2\cdot 10^{7}{\rm cm/s})^{2}}+1.2\cdot 10^{28}\,{\rm g}\cdot\frac{v_{0}^{4}}{\left(2.2\cdot 10^{7} {\rm
    cm/s}\right)^4}\quad\mbox{main seq.\ stars}\,,\\
m_{pbh}&>&6.6\cdot 10^{28}\,{\rm g}\cdot\frac{v_{0}^{2}}{(2.2\cdot 10^{7}{\rm cm/s})^{2}}+1.2\cdot 10^{27}\,{\rm g}\cdot\frac{v_{0}^{4}}{\left(2.2\cdot 10^{7} {\rm
    cm/s}\right)^4}\quad\mbox{RGCs}\,,\\
m_{pbh}&>&6.6\cdot 10^{29}\,{\rm g}\cdot \frac{v_{0}^{2}}{(2.2\cdot 10^{7}{\rm cm/s})^{2}}+1.2\cdot 10^{27}\,{\rm g}\cdot\frac{v_{0}^{4}}{\left(2.2\cdot 10^{7} {\rm
    cm/s}\right)^4}\quad\mbox{white dwarfs}\,,\\
m_{pbh}&>&6.6\cdot 10^{27}\,{\rm g}\cdot\frac{v_{0}^{2}}{(2.2\cdot 10^{7}{\rm cm/s})^{2}}+1.2\cdot 10^{22}\,{\rm g}\cdot\frac{v_{0}^{4}}{\left(2.2\cdot 10^{7} {\rm
    cm/s}\right)^4}\quad\mbox{neutron stars}\,.
\end{eqnarray}
On the view that the existence in the dark halo of PBHs with masses
greater than about $10^{26}$~g is already constrained by the
micro-lensing observations, we conclude that the capture of a PBH by a
star is difficult in {\it any} of the PBH mass ranges.
Only relative velocities far below the average value could lead to a
capture, but such events do not happen often enough to be significant. If
any PBHs are captured by stars, this must be extremely rare, and certainly
the fact that we do not observe them does not additionally constrain the
PBH abundance in the dark halo of our Galaxy\footnote{A few years ago,
Bohdan Paczynski suggested in a private conversation with M.A.A., that
some unusual and very rare supernova-like events might be due to PBH
collisions with white dwarfs. This suggestion was discussed later with
Piero Madau, Rashid Sunyaev and others, but never followed up in detail.}.

The above conclusion follows from the estimate of the energy lost when a
PBH passes through a star, $E_{\rm loss}$, based on the Ruderman-Spiegel
effective dynamical friction formula. This may explain the difference
between our conclusions and those of \cite{turolla2009}: they do not
estimate $E_{\rm loss}$, but instead assume that any PBH which collides
with a star would be captured.

We will examine the problem of estimating the probability
of very low velocity PBH-star collisions at high PBH masses in a separate publication.


\section{Conclusions \label{conclusion}}
 We have shown in the previous sections that the process of effective
friction when a PBH passes through a Galactic object cannot lead to a
significantly detectable signal with current instruments. We calculated
lower limits for the observation time for the best-suited detectors: XMM
Newton, Swift-BAT, and Fermi-GBM and Fermi-LAT. These lower limits were
based on the assumptions that
 \begin{itemize}
\item {\it The entire energy loss is radiated without significant delays
and at the wavelength at which it is produced,}
\item {\it one photon is sufficient for having a significant detection of
a signal, }
\item {\it all emitted energy is contained in the energy range of the
detector,}
\item {\it the entire thickness of the Galactic disk can be observed,}
\item {\it the signal is emitted at the largest opening angle, i.e.\ $2\pi$.}
\end{itemize}
 All of these assumptions were made in the sense of favoring detection;
more realistic settings would reduce the detectability even more. We have
investigated interactions with main sequence stars, red giants, white
dwarfs, and neutron stars. In general, we can conclude that the best results are obtained for main sequence stars.
However, independently of the class of objects, the
difficulty is that either the total energy loss in the process is too
small (for lower masses around $10^{15}$~g~$<m_{pbh}<10^{20}$~g) or the
event rate is too small (for larger masses around
$10^{20}$~g~$<m_{pbh}<10^{26}$~g). Even with the optimistic assumptions
listed above, observation times of more than $100$~years are required in
order to see a single event for all of the objects and all of the
detectors considered.

Although we use average values in place of the mass and velocity
distributions of PBHs, variations away from these central values will not
influence the result of these calculations significantly. One could, for
instance, argue that the clumping of dark matter as discussed by
\cite{binney_tremaine2008} would lead to an enhanced signal. However, such
an effect would need to increase the event rate by at least a factor of
$10^{2}$ in order to achieve reasonable observation times of
$t_{obs}<1$~yr. As shown in this paper, the central bulge of the
Galaxy may improve the situation by a factor of $20$, but this is
still not enough to reduce observation times to realistic scales of
less than a month.

Hence, we conclude that constraining the abundance of PBHs in DM using
interactions of PBHs with Galactic objects is not possible in the standard
scenario described above. A significant rate of GRB-like events as
suggested in \citep{zhilyaev2007} can therefore be excluded. There may be
additional aspects, such as clumping of dark matter or observation of nearby
dwarf galaxies, leading to an
enhanced event rate and energy release. However, these will have to be
able to improve the observation times in this standard scenario by at
least a factor of $100$ in order to give a significant signal. The main
reason is that observation times of less than a month may be available with the
different instruments, but not more.

Concerning a possible collision with the Earth: the predicted rate
is as low as $10^{-12}$\,yr$^{-1}\,\cdot\, m_{pbh,20}^{-1}$. This means
that there would be less than one such event per $10^{12\rightarrow
17}$~years for PBH masses above $10^{20}$~g. The interpretation of the
Tunguska event as a collision with a PBH of $m_{pbh}>10^{20}$~g, see
\cite{jackson_ryan1973}, can therefore be considered highly implausible on
the basis of this straightforward statistical argument. The most common
explanation discussed today is that a comet or asteroid disintegrated on
its way to Earth. Since no macroscopic fragments of the initial object
were found, its exact nature, comet or asteroid, is still unknown, see
e.g.\ \citep{jopek2008}.

On the other hand, PBHs with masses below $m_{pbh}<10^{17}$~g may have
struck or could still strike the Earth, as the encounter frequency may be
greater than once during the Earth's life time. However, the energy loss
is too small to see such events by means of radiation. The possibility of
detecting them by means of an acoustic signal has been discussed by
\cite{khriplovich2008} but there it was pointed out that the signal is too
weak to be seen by seismic detectors. It could be interesting to look for
those signatures with the planned generation of high-energy neutrino
telescopes, optimized to detect acoustic signals: large natural water, ice
or salt reservoirs are to be instrumented to look for acoustic signals
from Cherenkov radiation, see e.g. \citep{justin_madison06,julias_review}
for a review. The surface area of such a detector is planned to be of the
order of $10$~km$^{2}$. If the PBH has a distinct signature, it might be
observable with such an instrument.

\acknowledgments
 We would like to thank Wlodek Bednarek, Alina Donea, Maxim Khlopov, John Miller,
 Wolfgang Rhode and Alessandro Romeo  for helpful discussions. MAA acknowledges support from the
Polish Ministry of Science, grant N203 0093/1466, and from the Swedish
Research Council, grant VR Dnr 621-2006-3288. JKB is supported by the
Deutsche Forschungsgemeinschaft (DFG), grant BE-3712/3-1. Support for work
with PLB has come from the AUGER membership and theory grant 05 CU 5PD 1/2
via DESY/BMBF and from VIHKOS. AG acknowledges support from the Knut and
Alice Wallenberg Foundation. For LQ, support is coming from the Chinese
Scholarship Council. Further acknowledgments go to Nordita for traveling
grants for MAA, JKB and FJ.\\


\begin{deluxetable}{ll|lllll}
\tablecaption{Basic properties of objects considered for collisions with primordial
black holes.\label{object_properties}}
\startdata
\hline
&&Earth&main&red giant&white&neutron\\
&&&sequ.\ stars&cores&dwarfs&stars\\\hline
number in MW&$N_{\star}$&$1$&$10^{11}$&$10^{9}$&$10^{9}$&$10^{9}$\\
mass&$M_{\star}/M_{\odot}$&$3\cdot 10^{-6}$&1&1&1&1\\
radius&$R_{\star}$ [cm]&$6.4\cdot 10^{8}$&$10^{11}$&$10^{10}$&$10^{9}$&$10^{6}$\\
density&${\rho}_{\star}$ [g\,cm$^{-3}$]&$5.5$&$100$&$10^{4\rightarrow 5}$&$10^{5\rightarrow
  6}$&$10^{13\rightarrow 15}$\\
\enddata
\end{deluxetable}

\begin{deluxetable}{ll|llll}
\tablecaption{Basic dependence of the velocity $V(r)$, the mass density
  $\rho(r)$ and the mass $M_{tot}(r)$ on the distance from the Galactic
  Center $r$. Here, $M_{tot}(r)$ denotes the mass enclosed in the volume of
  the radius $r$. The parameters have different dependencies in the
  outer part of the Galaxy (``outside'') and the central part (``Bulge''). \label{table-regions}}
\startdata
\hline
~ & ~ & $V(r)$ & $\rho(r)$ &  $M_{tot}(<r)$ &
total enclosed mass\\ \hline
Bulge & $0 < r < r_D$ &
$\left(v_0/r_D\right)r$~ &
$\left(v_0^2\cdot r_D/G\right)$ &
$\left(v_0^2\cdot r_D/G\right)r^3$ & $v_0^2\cdot r_D/G$ \\
Outside & $r_D < r < r_{halo}$ & $\left(v_0\right)$ &
$\left(v_0^2/(4\pi G)\right)\cdot r^{-2}$ &
$\left(v_0^2/G\right)\cdot r$ & $v_0^2\cdot r_{halo}/G$

\enddata
\end{deluxetable}

\begin{deluxetable}{l|lllll}
\tablecaption{Energy loss as calculated in Section \ref{energy_deposit}, Equ.\ (\ref{eq:en_loss_num}) for a
PBH mass of
  $m_{pbh}=10^{20}$~g. The velocity of the PBH at the surface of the star $v_{\star}$ is calculated according to Equ.\ (\ref{vstar}). The energy loss rate, $dE_{\rm loss}/dt$ and the total energy loss $E_{\rm loss}$, scale as $(m_{pbh}/10^{20}\rm{
    g})^{2}$. The time it takes for the PBH to traverse the star is given by $dt$, see Section \ref{energy_deposit}. The peak energy $E_{\rm peak}$ is derived following Equ.\ (\ref{epeak_star}) in the case of regular stellar
objects, and using Equ.\ (\ref{epeak_rel}) for the relativistic objects (neutron stars and white dwarfs). The interaction probability $\dot{n}$ scales as
$(m_{pbh}/10^{20}\rm{
    g})^{-1}$ according to Equ.\ (\ref{ndot:equ2}). \label{energy:table}}
\startdata
\hline
$m_{pbh}=10^{20}$~g&Earth&main sequ.\ stars&red giant cores&white
dwarfs&neutron stars\\\hline
$v_{\star}$ [cm\,s$^{-1}$]&$2.2\cdot 10^{7}$&$5.5\cdot 10^{7}$&$1.6\cdot
10^{8}$&$5.1\cdot 10^{8}$&$1.6\cdot 10^{10}$\\
$dE_{\rm loss}/dt$ [erg\,s$^{-1}$]&$1.5\cdot 10^{21}$&$10^{22}$&$3\cdot
10^{23\rightarrow 24}$&$10^{24\rightarrow 25}$&$3\cdot 10^{30\rightarrow
32}$\\
$dt$ [s]&$60$&$3600$&$120$&$3.9$&$1.2\cdot 10^{-4}$\\
$dE_{\rm loss}/dt\cdot dt$ [erg]&$9\cdot 10^{22}$&$4\cdot 10^{25}$&$4\cdot
10^{25\rightarrow 26}$&$4\cdot
10^{24\rightarrow 25}$&$4\cdot 10^{26\rightarrow 28}$\\
$E_{\rm peak}$ [keV]&$0.2$&$1$&$11$&$1600$&$1.6\cdot 10^{6}$\\
$\dot{n}$ [yr$^{-1}$]&$8\cdot 10^{-13}$&$1.3\cdot 10^{4}$&$11$&$1.4$&$1.1\cdot
10^{-3}$
\enddata
\end{deluxetable}

\begin{deluxetable}{l||l|l|l|l}
\tablecaption{Detector properties in the X-ray to soft gamma-ray
  range. In order to give an upper limit on the sensitivity, the
  lower energy threshold is used as $E_{\gamma}$ in the case of Swift
  and Fermi. For XMM Newton, the reference energy $E_{\gamma}=1.5$~keV was
  explicitly given for the listed effective area.\label{detectors:table}}
\startdata
\hline
&XMM Newton&Swift BAT&Fermi GBM&Fermi LAT\\
&\citep{xmm_performance2002} &\citep{swift_web}
&\citep{fermi_gbm2008}&\citep{fermi_lat2003}\\\hline\hline
$E_{\min};\,E_{\max}$ [keV]&$(0.1;\,15)$&$(15;\,150)$&$(10;\,3\cdot 10^{4})$&$(2\cdot 10^{4};\,3\cdot 10^{8})$\\
$E_{\gamma}$ [keV]&$1.5$&$15$&$150$&$2\cdot 10^{4}$\\
$\Omega_{obs}$ [sr]&$0.20$&$2$&$\sim 2\pi$&$2$\\
$A_{eff}$ [cm$^{2}$]&2485&5200&$12\times 126$&$8\cdot 10^{3}$\\\hline\hline
main seq. stars&&&&\\
$m_{pbh}^{\rm break}$ [g]&$2.1\cdot 10^{24}$&$0.7\cdot 10^{24}$&$2.1\cdot 10^{25}$&$1.4\cdot 10^{26}$ \\
$t_{obs}^{\rm best}$ [yr]&$280$&$280$&$180$&$1500$\\\hline
Red Giant Cores&&&&\\
$m_{pbh}^{\rm break}$ [g]&$2.1\cdot 10^{24}$&$0.7\cdot 10^{24}$&$2.1\cdot 10^{25}$& $1.4\cdot 10^{26}$\\
$t_{obs}^{\rm best}$ [yr]&$2.8\cdot 10^{5}$&$2.8\cdot 10^{5}$&$1.2\cdot 10^{5}$&$1.5\cdot 10^{6}$\\\hline
White Dwarfs&&&&\\
$m_{pbh}^{\rm break}$ [g]&$4.2\cdot 10^{24}$&$2.1\cdot 10^{24}$&$4.2\cdot 10^{25}$&$2.8\cdot 10^{26}$ \\
$t_{obs}^{\rm best}$ [yr]&$1.0\cdot 10^{7}$&$1.1\cdot 10^{7}$&$4.8\cdot 10^{6}$&$7.6\cdot 10^{7}$\\\hline
Neutron Stars&&&&\\
$m_{pbh}^{\rm break}$ [g]&$2.1\cdot 10^{26}$&$7.1\cdot 10^{25}$&$2.1\cdot 10^{27}$&$1.4\cdot 10^{28}$\\
$t_{obs}^{\rm best}$ [yr]&$2.8\cdot 10^{4}$&$2.8\cdot 10^{4}$&$1.2\cdot 10^{4}$&$1.5\cdot 10^{7}$\\\hline
\enddata
\end{deluxetable}
\end{document}